# Gate-tunable magnetoresistance in six-septuple-layer MnBi$_2$Te$_4$

Yaoxin Li[1†], Chang Liu[1,2†], Yongchao Wang[3†], Hao Li[4,5], Yang Wu[5,6],

Jinsong Zhang[1,7*], Yayu Wang[1,7*]

[1]*State Key Laboratory of Low Dimensional Quantum Physics, Department of Physics, Tsinghua University, Beijing 100084, P. R. China*

[2]*Beijing Academy of Quantum Information Sciences, Beijing 100193, P. R. China*

[3]*Beijing Innovation Center for Future Chips, Tsinghua University, Beijing 100084, P. R. China*

[4]*School of Materials Science and Engineering, Tsinghua University, Beijing, 100084, P. R. China*

[5]*Tsinghua-Foxconn Nanotechnology Research Center, Department of Physics, Tsinghua University, Beijing 100084, P. R. China*

[6]*Department of Mechanical Engineering, Tsinghua University, Beijing 100084, P. R. China*

[7]*Frontier Science Center for Quantum Information, Beijing 100084, P. R. China*

[†] *These authors contributed equally to this work.*

\* Emails: jinsongzhang@tsinghua.edu.cn; yayuwang@tsinghua.edu.cn

**Abstract:** The recently discovered antiferromagnetic topological insulator MnBi$_2$Te$_4$ hosts a variety of exotic topological quantum phases such as the axion insulator and Chern insulator states. Here we report systematic gate voltage dependent magneto transport studies in six septuple-layer MnBi$_2$Te$_4$. In *p*-type carrier regime, we observe positive linear magnetoresistance when MnBi$_2$Te$_4$ is polarized in the ferromagnetic state by an out-of-plane magnetic field. Whereas in *n*-type regime, distinct negative magnetoresistance behaviors are observed. The magnetoresistance in both regimes is highly robust against temperature even up to the Néel temperature. Within the antiferromagnetic state, the behavior of magnetoresistance exhibits a transition from negative to positive as applying a gate voltage. The boundaries between different magnetoresistance behaviors in the experimental phase diagram can be explicitly characterized by the gate-voltage-independent magnetic fields that denotes the processes of the spin-flop transition. The rich transport phenomena demonstrate the intricate interplay between topology, magnetism and dimensionality in MnBi$_2$Te$_4$.

## 1. Introduction

Topological quantum materials such as topological insulators (TIs) [1,2] and Weyl semimetals (WSMs) [3–5] can produce a variety of exotic topological quantum phenomena when the nontrivial topological band structure interplays with other degrees of freedom. An outstanding example is the quantum anomalous Hall (QAH) effect that was first realized in magnetically doped TI films with ferromagnetic (FM) order, which is now called the Chern insulator phase [6–11] due to the absence of Landau levels. The axion insulator is another topological state defined in a bulk insulator that is characterized with nontrivial axion field $\theta$ [12–16]. Early experimental demonstrations of an axion insulator are mainly built on TI heterostructures with the top and bottom surface states gapped by opposite magnetization [17–19]. By applying an out-of-plane magnetic field, the axion insulator can be driven into the QAH insulator phase when the two surfaces are polarized in the FM state.

Recently, the intrinsic magnetic TI $MnBi_2Te_4$ has drawn a lot of attentions in the topological state of matter community. It is found to be a TI with A-type antiferromagnetic (AFM) order below the Néel temperature $T_N \sim 25$ K [20–26] at zero magnetic field. A moderate magnetic field of about 5 T can drive it to a FM WSM [27]. When it is exfoliated into even-septuple-layer (even-SL) such as 6-SL, it is proposed and demonstrated to exhibit the axion insulator state in AFM regime at low magnetic field and the Chern insulator state in FM regime at high magnetic field [21,22]. However, previous works on 6-SL $MnBi_2Te_4$ mainly focus on the quantum phenomena with Fermi level ($E_F$) tuned near the charge-neutral point (CNP). The gate voltage ($V_g$) dependence of magneto transport and the interplay between band topology and magnetic order in conductive regimes have not been thoroughly investigated. Although the MR behavior in $MnBi_2Te_4$ has been reported by some recent works [28,29], a detailed investigation of the $V_g$ dependent transport behaviors has been lacking due to the absence of quantization and insufficient gate tunability on thick samples. In addition, the WSM nature of FM $MnBi_2Te_4$ may further give rise to exotic topological quantum phenomena as demonstrated in other WSM systems, such as unsaturated linear magnetoresistance (MR) [30–32] and chiral anomaly [30,33,34].

In this work, we fabricate high-quality 6-SL $MnBi_2Te_4$ device and measure the MR and Hall traces at different $V_g$s and temperatures ($T$s). We observe the axion insulator state to Chern insulator state transition in the most insulating regime with $E_F$ in the CNP. Then we thoroughly investigate the magneto transport behaviors in the 6-SL device. In the FM state, we find a positive linear MR characterizing the nature of WSMs persists up to $T_N$ where the Hall traces indicate *p*-type carriers. In contrast, as the transport is tuned to the *n*-type regime, the slope of linear MR turns to negative. While in the low-field AFM state, we observe a reversal from negative to positive MR along with the transition from *p*-type to *n*-type regime at $T < T_N$. The experimental phase diagram summarized from the $V_g$ dependent MR is consistent with the origin of magnetic order in $MnBi_2Te_4$.

## 2. Methods

$MnBi_2Te_4$ single crystals are grown by direct reaction of $Bi_2Te_3$ and MnTe with the ratio of 1:1 in a vacuum-sealed silica ampoule. The mixture is heated to 973 K and then cooled down to 864 K slowly. The quality of $MnBi_2Te_4$ crystal and its precursors is examined by X-ray diffraction, EDX spectra and high-resolution XPS spectra. The growth method and characterization are similar to that in previous works [25,26]. The size of the $MnBi_2Te_4$ bulk crystal in use is about 2 mm. Thin flakes are exfoliated onto 285-nm-thick $SiO_2$/Si substrates treated by air plasma. The thickness of 6-SL is determined by the optical contrast as commonly used in previous studies and checked by atom force microscope measurements (Supplementary Fig. S1). The Hall-bar structure is fabricated by electron beam lithography followed by thermal evaporation of Cr/Au. All the device fabrication processes are carried out in argon-filled glove box. Before transport measurement, the devices are covered with 400 nm thick PMMA. Low-temperature transport measurements are performed in a commercial cryostat with the base temperature of 1.6 K and magnetic field up to 9 T by using the standard four-probe lock-in techniques with low-frequency (12.357 Hz) and ac current of 200 nA. Both the longitudinal resistance ($R_{xx}$) and Hall resistance ($R_{yx}$) are measured simultaneously using a pair of SR830 amplifiers. $V_g$ is applied to the sample by a Keithley 2400 voltage source.

## 3. Results and discussion

The magnetic and crystal structures of 6-SL $MnBi_2Te_4$ are shown in Fig. 1(a). At zero magnetic field, the $Mn^{2+}$ moments between neighboring SLs are antiferromagnetically aligned. As an out-of-plane magnetic field is applied, $MnBi_2Te_4$ is polarized to the FM state. We fabricate 6-SL $MnBi_2Te_4$ into a standard 6-probe Hall bar geometry and firstly measure the $R_{xx}$ and $R_{yx}$ at different $V_g$s at $T$ = 1.6 K. Figure 1(b) and 1(c) display the variations of $R_{xx}$ and $R_{yx}$ with $V_g$ at $\mu_0 H$ = 0 T and 9 T respectively. At $\mu_0 H$ = 0 T, a typical insulating behavior shows up at the CNP around $V_g$ = 30 V as manifested by a large $R_{xx}$ over 150 kΩ. In the $V_g$ range from 24 V to 40 V, $E_F$ is tuned towards the band gap and the system enters the axion insulator state regime. When a magnetic field is applied, MnBi$_2$Te$_4$ is driven into the FM Chern insulator state, characterized with a quantized $R_{yx}$ plateau at $h/e^2$ and vanishing $R_{xx}$. The magnetic-field-driven axion insulator to Chern insulator topological phase transition is displayed in the fourth panel of Fig. 1(d). All these transport characters are hihgly consistent with previous reports [25,35–37].

Then we focus on the magneto transport in the regimes away from the CNP. Figure 1(d) shows the representative behaviors of MR and Hall traces at different $V_g$s. For $V_g \leq 12$ V, $R_{yx}$ shows overall p-type behavior in both AFM and FM state, indicating that $E_F$ lies in the valence band. In contrast, $R_{yx}$ shows the overall n-type behavior and $E_F$ lies in the conduction band for $V_g \geq 50$ V. All Hall traces show two characteristic magnetic fields $H_{c1} \sim 2.3$ T and $H_{c2} \sim 5.2$ T, which correspond to the beginning and ending of spin-flop process respectively, as labelled in Fig. 1(d). These features can be also identified in the MR curves. Notably, the $R_{xx}$ curves show systematic and complex features with varied magnetic field. In the FM state, $R_{xx}$ shows a positive linear MR in the p-type regime. While in the AFM state, the negative MR in the p-type regime evolves into a positive one in the n-type regime. The distinct behaviors in MR reflects the close relationship between the carrier type and magnetic structure, which is the main focus of this work.

To understand these MR behaviors, we perform magneto transport on 6-SL MnBi$_2$Te$_4$ at varied $T$s. In Fig. 2(a) and 2(b) we compare the $T$ dependent MR curves for $V_g$ = 0 V and 50 V. A magnetic field independent $R_{xx}$ is observed in the AFM state for $V_g$ = 0 V, except for a weak peak at ~0.2 T (see Supplementary Fig. S2 for details). The positive linear MR in the FM state persists up to 25 K which is above $T_N$ of 6-SL MnBi$_2$Te$_4$. As $T$ is increased, $H_{c1}$ and $H_{c2}$ gradually decrease and drop to zero for $T \geq 20$ K so that the field range of linear MR increase systematically. In contrast to the positive MR at 0 V, the MR at 50 V in the FM state becomes negative. With the increase of $T$, $R_{xx}$ in the FM state gradually increases and the negative MR persists to above $T_N$.

In previous studies, robust linear MR against magnetic field and temperature has been reported in many topological materials such as TIs [38–41] and WSMs [30–32,42,43]. And the WSM nature of the FM state of $MnBi_2Te_4$ is suggestive of the possible similar origins [21,22,36,44]. In previous works, the explanations of positive linear MR in WSMs are mainly focused on the classical [45,46] or quantum linear MR theory [47,48]. Quantum theory predicts that MR should be proportional to $1/n^2$, where $n$ denotes the carrier density. According to this model, linear MR occurs in the extreme quantum limit when only few Landau levels are occupied. Although the hole density is very low for $V_g = 0$ V due to the linear dispersion of the WSM state of $MnBi_2Te_4$ [~$10^{12}$ cm$^{-2}$, shown in Fig. 2(c)], the magnetic field at which we observe linear MR is far from the criterion of quantum limit $n << (eH/c\hbar)^{3/2}$ [48]. Furthermore, as plotted in Fig. 2(c), carrier density against $dMR/\mu_0 dH$ curve at different $T$s strongly deviates from the $1/n^2$ dependence predicted by the theoretical model. Here MR is expressed as $(R_{xx}(\mu_0 H) - R_{xx}^0) / R_{xx}^0$, where $R_{xx}^0$ is the intercept of the linear fitting of $R_{xx}$ - $\mu_0 H$ in the FM state for $V_g = 0$ V and $R_{xx}$ at 0 T for other $V_g$s. Therefore, the quantum theory fails to explain the linear MR here. On the other hand, the classical linear MR theory considering the mobility fluctuation predicts that the MR is proportional to the larger of average mobility ($\mu$) and the variance of mobility ($\Delta\mu$) [45]. In $MnBi_2Te_4$, it has been reported by many works that $\mu$ is very low due to the prevalence of vacancies and anti-site defects [24,25,49,50]. When the carrier density is too low to screen the electric potential fluctuation induced by disorders, linear MR will occur [45]. As Fig. 2(c) shows, $dMR/\mu_0 dH$ indeed increases with increasing mobility $\mu$ though it slightly deviates from linear dependence. It is worth noting that the $\mu$ we obtain from transport data is the average mobility and the linear relation of $\mu$ only satisfies when the value of $\mu$ is larger than $\Delta\mu$. Large fluctuation of mobility in $MnBi_2Te_4$ may explain the deviation from linear dependence. But further experiments about the microscopic distribution of defect are required to verify the model completely.

For $V_g = 50$ V, $E_F$ is tuned out of the band gap and crosses the conduction band. However, the overall $R_{xx}$ curve at $V_g = 50$ V is qualitatively the same as that at the CNP with $V_g = 30$ V except for the quantized transport. In the Chern insulator state of 6-SL $MnBi_2Te_4$ at the CNP, $E_F$ is within the band gap and intersects with the dissipationless chiral edge state. Magnetic field can localize other dissipative conduction channels generated by thermal excitations and the negative MR occurs. The calculated band structure of few-layer $MnBi_2Te_4$ in the FM state shows that the chiral edge state survives even when $E_F$ lies beyond the band gap and coexists with the conduction band for a large energy range [27,36,51]. The considerable scattering between the chiral edge state and bulk state hiders the perfect quantization but the negative MR induced by the same mechanism can still occur. The Weyl band induced positive MR is drowned by the chiral edge conduction, which is absent is the *p*-type regime. Asymmetric electron and hole bands in FM state can explain the sign change in MR at different $V_g$s. And the metallic $R_{xx}$ vs. *T* relation in the FM state for $V_g = 50$ V is also consistent with this scenario.

Now we turn to the MR behavior in the AFM state. As Fig. 1(d) shows, when we set $V_g = 12$ V, it is still in the *p*-type regime. The positive linear MR in the FM state persists but a distinct negative linear MR in the AFM state emerges, which differs from that at 0 V. A MR peak at ~ 0.2 T is also observed. At higher *T*s, as shown in Fig. 3(a), the slope of MR decreases and the field range of negative MR shrinks. At $T = 20$ K and 25 K, the negative MR at low magnetic field vanishes and turns to positive in the whole field range. For clarity we only plot MR curves between ±3.5 T and the complete data from -9 T to 9 T is shown in Supplementary Fig. S5. Interestingly, the MR curves for $V_g \geq 30$ V show the opposite behavior. At low *T*s, a positive MR is observed in the AFM state and the MR peak at $V_g = 12$ V turns into a dip. The slope of MR increases with increasing temperature up to 16 K, as shown in Fig. 3(b). At 20 K and 25 K, however, the slope of MR suddenly becomes negative. In Fig. 3(c), we display the temperature dependent $dMR/\mu_0 dH$. When $T > T_N$, the AFM order vanishes and the MR shows the behavior like that in the FM state as discussed above.

At lower *T*s, the sign of MR and MR peak/dip are gate dependent, behaving like that in Mn-doped $Bi_2Te_{3-y}Se_y$ [52]. The butterfly-shaped MR in the *p*-type regime is the conventional behavior associated with the suppression of spin scattering by magnetic field [24,53]. And the linear MR seems like that in the system hosting electron-magnon scattering [54–58]. When the temperature is raised to above $T_N$, the direction of local moments is random so that the spin-wave is absent and the magnon related negative MR vanishes. However, this behavior only emerges in a narrow gate range around $V_g = 12$ V. The study on the excitation of spin-wave and its gate dependence in $MnBi_2Te_4$ is desired to verify the scenario. When the $E_F$ is tuned towards the CNP, magnetic field tends to suppress the conductivity in the AFM state, turning the MR into positive. We attribute this behavior to the emergence of a pair of helical hinge conduction in the axion insulator state, which is closely related to the AFM magnetic structure [59,60]. When magnetic field breaks the $S$ symmetry in the axion insulator state, the axion field $\theta$ deviates from the quantized value and the scattering between the helical conduction channels is enhanced. Similar MR behavior has been observed in many QSH systems that host helical edge state protected by time-reversal symmetry [61,62]. The competing mechanism between magnon induced negative MR and edge conduction induced positive MR result in the temperature dependence of the slope of MR. The change in sign of MR at different regions can be also interpreted by this competing mechanism.

In Fig. 4(a), we depict the color map of $dMR/\mu_0 dH$ in $\mu_0 H$-$V_g$ plane at 1.6 K. There are four prominent regions I~IV with two clear phase boundaries $H_{c1}$ and $H_{c2}$. Region I and III are located in the $p$-type regime and II and IV cross the charge-neutral regime and the $n$-type regime. Below $H_{c1}$, $dMR/\mu_0 dH$ changes from a negative value in region III to a positive one in region IV due to the competition mechanism between electron-magnon scattering and edge conduction in the axion insulator state as discussed above. When the magnetic field is tuned to above $H_{c2}$, the Weyl band induced positive MR in region I changes into negative MR due to the emergence of the chiral edge state in region II. Two characteristic magnetic fields $H_{c1}$ and $H_{c2}$ at different $V_g$s are both nearly constant. The carrier independent characteristic fields are consistent with the origin of magnetism in MnBi$_2$Te$_4$, which is mediated by the superexchange interaction between Mn-Te-Mn [21,22]. When the temperature is increased, $H_{c1}$ and $H_{c2}$ decrease and near vanish above 20 K, the $T_N$ of 6-SL MnBi$_2$Te$_4$, as shown in Fig. 4(b). We only plot the color map of $dMR/\mu_0 dH$ at $V_g = 30$ V here because the temperature evolutions of $H_{c1}$ and $H_{c2}$ are gate independent (see Supplementary Fig. S4). The distinct AFM regime with a positive MR and FM regime with a weak negative MR are separated by the spin-flop regime at low $T$s and merge into paramagnetic regime above 20 K.

## 4. Conclusions

The gate dependent magneto transport of 6-SL MnBi$_2$Te$_4$ device displays systematic and complex behaviors resulting from the interplay among band topology, magnetism and scattering mechanism. Distinct MR behaviors are observed in different parameter ranges. In the AFM state, we observe negative MR in the $p$-type regime, while it becomes positive when $E_F$ is tuned from the CNP to the conduction band. This behavior is explained by the competing between spin scattering mechanism and edge conduction in the axion insulator state. In the FM state, owing to the asymmetric band structure, we observe a reversal from positive linear MR in the *p*-type regime to a negative MR in the *n*-type regime. The characteristic magnetic fields labeling the boundary of the AFM, spin-flop progress and the FM state are independent of $V_g$. When the temperature is increased to above $T_N$, AFM order disappears and all the critical magnetic fields are reduced to zero. With the presence of such rich MR phenomena, MnBi$_2$Te$_4$ serves as an ideal platform for exploring the interplay between band topology and intrinsic magnetism.

**FIGURE CAPTIONS**

**Figure 1.** (a) Crystal and magnetic structure of 6-SL MnBi$_2$Te$_4$ in the AFM and FM states. (b)-(c) Gate dependent $R_{xx}$ and $R_{yx}$ at $T = 1.6$ K at (b) $\mu_0 H = 0$ T and (c) $\mu_0 H = 9$ T. The CNP is about 30 V. (d) MR and Hall traces at different $V_g$s. Navy and magenta arrows mark the characteristic magnetic fields $H_{c1}$ and $H_{c2}$ respectively. Navy and magenta broken lines demote the quantized Hall and vanishing longitudinal resistance plateau respectively.

**Figure 2.** (a)-(b) Magnetic-field dependence of $R_{xx}$ at different $T$s at 0 V and 50 V. (c) Mobility and hole carrier density plotted against d$MR$/$\mu_0$d$H$ in the FM state at different $T$s for $V_g = 0$ V.

**Figure 3.** (a)-(b) Magnetic-field dependence of MR at different $T$s for $V_g = 12$ and 50 V. (c) Temperature dependent d$MR$/$\mu_0$d$H$ between 1 T and 2T in the AFM state for $V_g = 12$ and 50 V. Magenta broken line denotes Néel temperature $T_N$.

**Figure 4.** (a) Color map of d$MR$/$\mu_0$d$H$ at $T = 1.6$ K in $\mu_0 H$-$V_g$ plane. Region I~IV labeled in diagram represent the typical MR behaviors discussed in text. $H_{c1}$ and $H_{c2}$ from the data in Fig. 1(d) plotted in navy and magenta broken lines demote the phase boundary. Black broken line marks the CNP with $V_g = 30$ V. (b) Phase diagram of d$MR$/$\mu_0$d$H$ at $V_g = 30$ V in $T$-$\mu_0 H$ plane. AFM, spin-flop and FM state are separated by the navy and magenta broken lines which show the temperature

dependent $H_{c1}$ and $H_{c2}$.

**Data Availability:** All raw and derived data used to support the findings of this work are available from the authors on request.

**Acknowledgements:** This work is supported by the Basic Science Center Project of NSFC (grant No. 51788104), the National Key R&D Program of China grants No. 2018YFA0307100and NSFC grant No. 51991340 and No. 21975140. This work is supported in part by Beijing Advanced Innovation Center for Future Chip (ICFC).

**Author contributions:** Y. Y. W. and J. S. Z. supervised the research. Y. X. L, C. L. and Y. C. W. fabricated the devices and performed the transport measurements. H. L. and Y. W. grew the MnBi$_2$Te$_4$ crystals. J. S. Z, Y. Y. W., Y. X. L. and C. L. prepared the manuscript with comments from all authors.

**Figure 1**

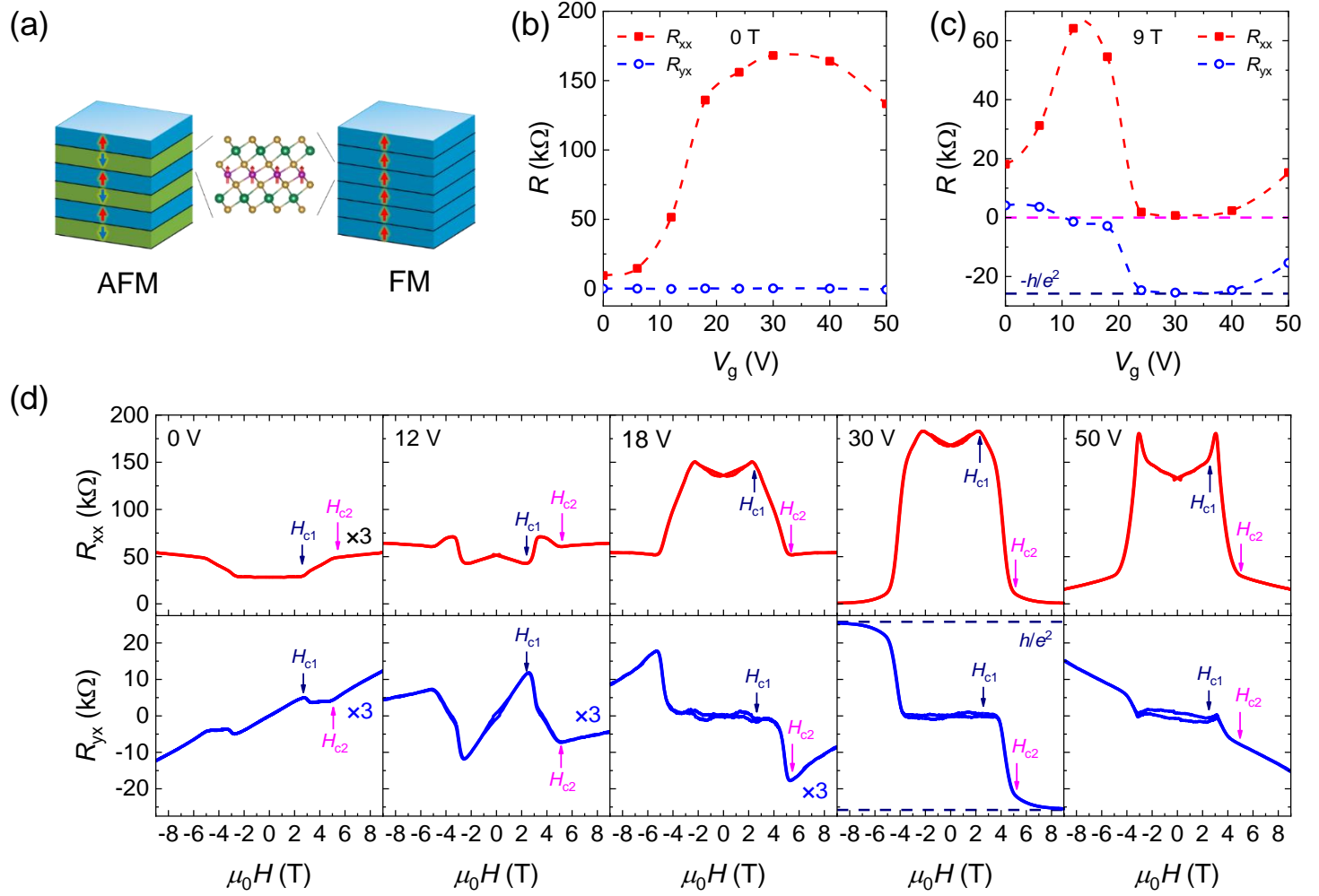

**Figure 2**

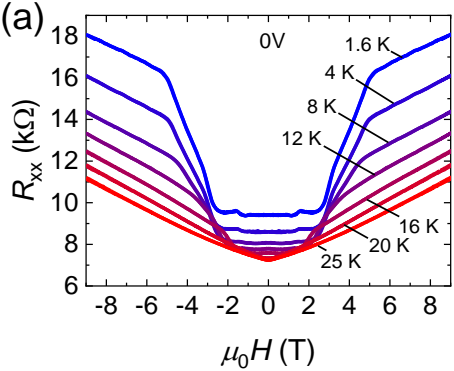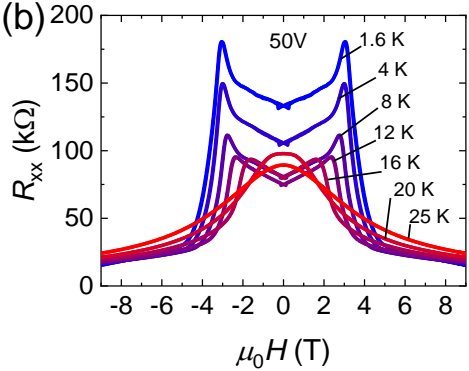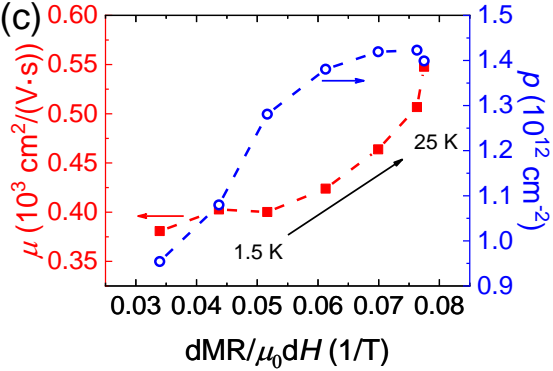

**Figure 3**

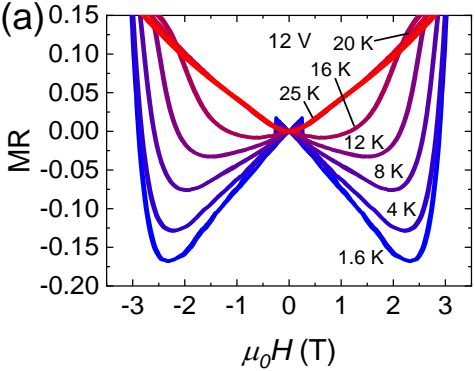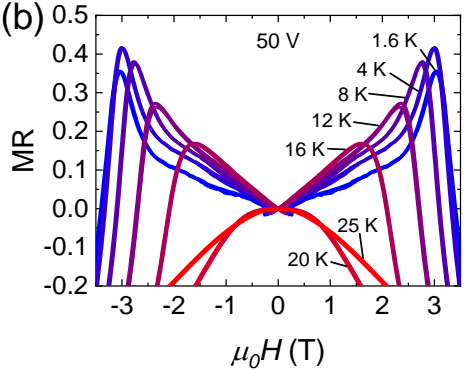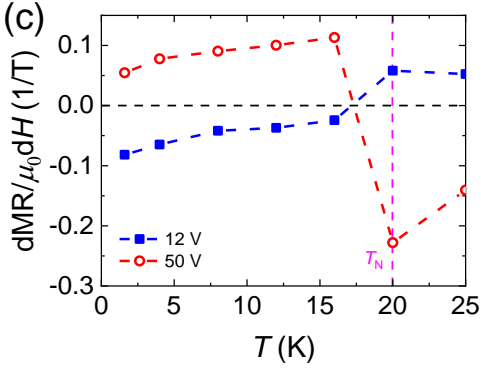

**Figure 4**

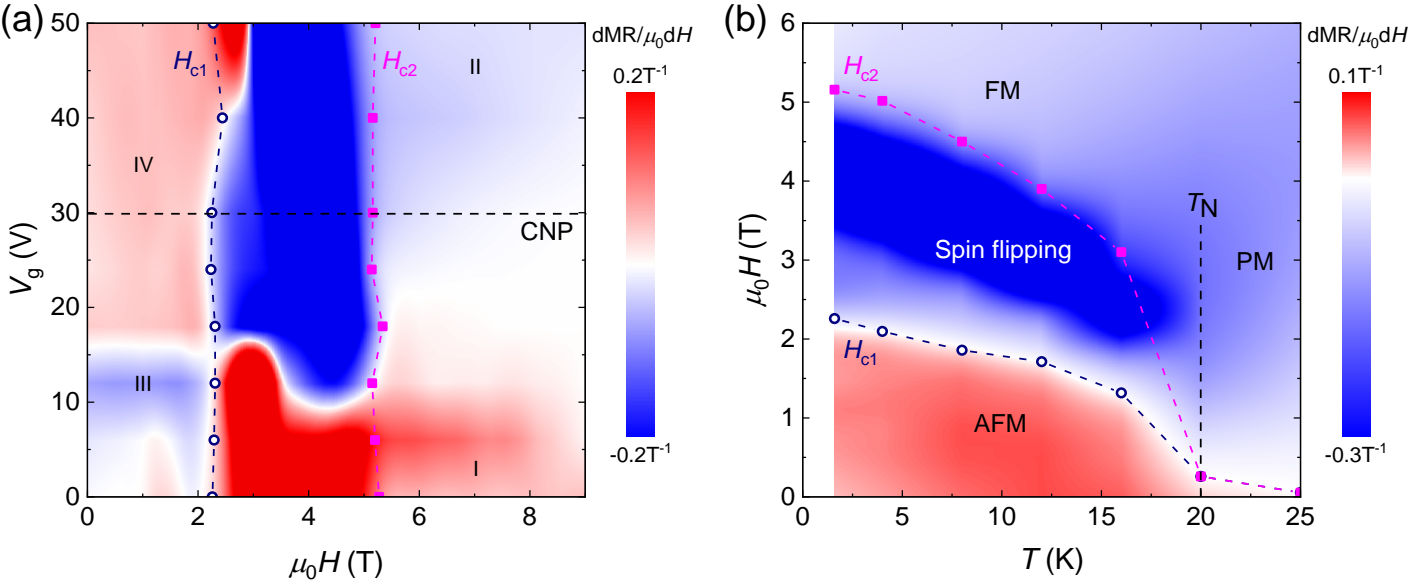